\documentclass[12pt]{article}
\usepackage{graphicx}
\usepackage{amssymb}
\usepackage{amsmath}
\usepackage{color}
\usepackage{here}

\begin{document}

\newcommand{\be}{\begin{equation}}
\newcommand{\ee}{\end{equation}}
\newcommand{\cV}{\mathcal{V}}
\newcommand{\cR}{\mathcal{R}}
\newcommand{\cP}{\mathcal{P}}
\newcommand{\xv}{\vec{x}}
\newcommand{\kv}{\vec{k}}

\baselineskip 0.65cm
\begin{titlepage}

\begin{flushright}
ICRR-Report-574-2010-7 \\
IPMU10-0185 \\
\end{flushright}

\vskip 1.35cm
\begin{center}


{\Large\bf
K\"{a}hler moduli double inflation
}

\vskip 1.5cm

{\large Masahiro Kawasaki$^{(a, b)}$, Koichi Miyamoto$^{(a)}$}

\vskip 0.4cm

{\it $^a$Institute for Cosmic Ray Research,
     University of Tokyo, Kashiwa, Chiba 277-8582, Japan\\
$^b$Institute for the Physics and Mathematics of the Universe, 
     University of Tokyo, Kashiwa, Chiba 277-8568, Japan
}


\vskip 1.5cm

\begin{abstract}
We show that double inflation is naturally realized
in K\"{a}hler moduli inflation, which is caused by 
moduli associated with string compactification.
We  find that there is a small coupling between the two 
inflatons which leads to amplification of perturbations 
through parametric resonance in the intermediate stage of double 
inflation.
This results in the appearance of a peak in the power spectrum of 
the primordial curvature perturbation. 
We numerically calculate the power spectrum and show that the power
spectrum can have a peak on observationally interesing scales. 
We also compute the TT-spectrum of CMB based on the power spectrum 
with a peak and see that it better fits WMAP 7-years data.
\end{abstract}

\end{center}
\end{titlepage}

\setcounter{page}{2} 

\section{Introduction}
 
The inflation is now a standard paradigm of the early universe.
It not only solves many problems associated with the early universe 
but also explains the origin of the density perturbation of the universe.
That is, the inflation generates the density perturbation from 
the quantum fluctuation of inflaton, the scalar field 
whose potential energy drives exponential expansion of the universe.
Search for the inflationary model which is based on particle physics 
has become one of the main topics of cosmology.
In particular, recently, inflation models realized 
in framework of superstring 
theory are studied vigorously.
Superstring theory predicts the existence of many scalar fields, 
moduli, in the low-energy effective theory and some of them can be 
candidates of inflaton(See~\cite{HenryTye:2006uv,Cline:2006hu,
Burgess:2007pz,McAllister:2007bg,Baumann:2009ni} 
for comprehensive reviews) .
 
Vacuum expectation values (VEVs) of some of moduli determine 
the geometry of the extra dimensions.
It is believed that in the vacuum all moduli are stabilized 
at the minimum of their potential and the extra dimensions are 
compactified to a specific Calabi-Yau manifold.
Comprehension of how to stabilize moduli, especially in the context 
of type-I\hspace{-.1em}IB superstring, has greatly advanced 
in this decade~\cite{Giddings:2001yu,Kachru:2003aw} 
(see also \cite{Grana:2005jc} for a review and references therein).
Following the stabilization mechanism, we can derive the potential 
of moduli.
It was found that some of moduli may be apart from their minimum 
at first and slowly roll down toward the minimum along the potential 
in the early universe, which leads to the sufficiently long 
inflation~\cite{BlancoPillado:2004ns,BlancoPillado:2006he,
Conlon:2005jm,Bond:2006nc,Cicoli:2008gp}. 
Among inflation models driven by various moduli, here, 
we concentrate on K\"{a}hler moduli inflation, 
which was first considered in~\cite{Conlon:2005jm}
(see elso \cite{BlancoPillado:2009nw}).
 
Each K\"ahler modulus, $T_i=\tau_i+i\theta_i$ is associated 
with a 4-cycle on the Calabi-Yau manifold.
The real part $\tau_i$ represents the volume of the 4-cycle 
and the imaginary part $\theta_i$ is the axionic component.
K\"ahler moduli inflation is based on the Large-volume flux 
compactification~\cite{Balasubramanian:2005zx,Conlon:2005ki}, 
which is a different compactification mechanism from KKLT 
scenario~\cite{Kachru:2003aw}.
In this mechanism, the total volume of the extra dimension is 
exponentially large, and the K\"ahler modulus with largest 
real part is responsible for the total volume.
The real part of another K\"{a}hler modulus may become an inflaton.
  
For general Calabi-Yau manifolds, the number of the K\"{a}hler 
moduli is more than one.
In fact, the K\"{a}hler moduli inflation requires at least three 
K\"{a}hler moduli.
The first one corresponds to the total volume, the second one is 
needed to stabilize the first one and the third one is an inflaton.
However, there is no reason that only three K\"ahler moduli
exist and we expect more than three K\"{a}hler moduli.
The models in which the total number of K\"{a}hler moduli is four 
and two of them behave as inflatons were 
considered in~\cite{Yang:2008ns,Berglund:2009uf}.
Therefore, in this paper, we consider such situation 
that two K\"{a}hler moduli play a role of inflatons and hence
two stages of the inflation take place.
Especially, we assume that the duration of the second inflation 
is not long enough to erase the effect of the first inflation. 
Such a two-stage inflationary model driven by two scalar fields 
is called double inflation~\cite{Starobinsky:1986fxa,Kofman:1986wm,
Silk:1986vc}.
In the simplest double inflation,  each inflaton has 
no(or only weak) coupling to another, 
so that the dynamics of the first inflaton  
does not affect the movement of second one.
In other words, the potential of inflatons must has a form like
\be
  V(\phi_1,\phi_2)=V_1(\phi_1)+V_2(\phi_2). \label{doubleinfVform}
\ee
This form is naturally realized for K\"{a}hler moduli.
This, in addition to the plethora of moduli, means that the double 
inflation, or the inflation which consists of more than two stages 
is natural rather than possible in superstring theory.
  
In this paper, we investigate the power spectrum of the density 
perturbation generated during the double inflation caused by 
two K\"{a}hler moduli.
The perturbations produced by various models of the double inflation 
were studied in many papers~\cite{Starobinsky:1986fxa,Kofman:1986wm,
Silk:1986vc,Salopek:1988qh,Polarski:1992dq,Langlois:1999dw,
Kanazawa:1999ag,Lesgourgues:1999uc,Gordon:2000hv,Tsujikawa:2002qx,
Kawasaki:2006zv}.
The power spectrum of the density perturbation from the double inflation 
has features which do not exist in that from the single-field inflation.
The Fourier modes of the perturbation which exit Hubble horizon 
during the first inflation have different amplitude and tilt from 
those which exit horizon during the second inflation.
Besides, as pointed out in~\cite{Kawasaki:2006zv}, the power 
spectrum might have a characteristic peak due to the parametric 
resonance~\cite{Kofman:1994rk,Shtanov:1994ce,Kofman:1997yn}.
During the intermediate stage between two inflationary stages, 
the first inflaton oscillates around the minimum of its potential, 
and its fluctuation  can grow exponentially through parametric 
resonance.
Then, if the first inflaton has a small coupling to the second one, 
forced oscillation occurs and the fluctuation of the second inflaton 
also grows exponentially.
After the parametric resonance becomes inefficient, the amplified 
fluctuation of the first inflaton decreases by the cosmic expansion. 
On the other hand, since the second inflaton is light during the second stage, its fluctuation remains amplified.
Therefore, the curvature perturbation is also amplified and its Fourier 
modes which exit the horizon with large amplitude sustain it 
outside the horizon.
Modes affected by this resonance effect are only those which exist 
in the resonance band while the first inflaton oscillates with large 
amplitude.
Then, the resultant power spectrum has a sharp peak.
We perform numerical integration of the system of the equations which describe the evolution of various perturbation variables 
during the inflation, and compute the power spectrum generated 
by the double inflation by two K\"{a}hler moduli.
We show that in this model the power spectrum has a peak of the type 
explained above.
  
Next, as one of implications of the double moduli inflation, 
we discuss the influence of this peak on the spectrum of the temperature 
fluctuation of the cosmic microwave background(CMB).
It is known that there is an upward deviation of the observational 
value compared to the 
predicted value around $\ell \simeq 40$~\cite{Larson:2010gs}.
We attribute this deviation to the existence of the small peak in the 
inflationary power spectrum produced by the above mechanism, and 
show that the TT spectrum fits better around $\ell \simeq 40$.
 
This paper is organized as follows.
In Sec.~\ref{sec:Kahler_moduli}, we review K\"{a}hler moduli inflation.
We present the equations which describe the evolution 
of the perturbations and the formalism to compute the power spectrum
in Sec.~\ref{sec:perturbation_formalism}.
In Sec.~\ref{sec:powerspectrum}, we numerically integrate 
the system of the equation in Sec.~\ref{sec:perturbation_formalism} 
and obtain the power spectrum.
Here, we see that it actually has a peak.
In Sec.~\ref{sec:TTspectrum}, we compute the TT spectrum of CMB 
perturbation using the power spectrum obtained in 
Sec.~\ref{sec:powerspectrum} and fit the theoretical spectrum to the observational data, especially around  $\ell \simeq 40$.
We summarize this paper in Sec.~\ref{sec:summary}.
 
Throughout this paper, we use the Planck unit, i.e.  $M_p=1$.

\section{Review of K\"{a}hler moduli inflation}
\label{sec:Kahler_moduli}
 
In this section, let us  briefly review K\"{a}hler moduli 
inflation~\cite{Conlon:2005jm,BlancoPillado:2009nw}.
K\"{a}hler moduli inflation is based on I\hspace{-.1em}IB flux 
compactification.
In this framework, as a low-energy effective theory, we can consider 
$4$-dimensional $\mathcal{N}=1$ supergravity characterized by 
the following superpotential and K\"{a}hler potential 
in the string frame:
\be
 W=W_0+\sum_{i=2}^{h^{1,2}} A_ie^{-a_iT_i} \label{W}
\ee
\be
 K=K_{cs}-2\ln\left(\cV+\frac{\xi}{2}\right). \label{K}
\ee
Here, $W_0$ is the superpotential for complex structure moduli 
and dilaton, which is induced by the background flux on 
the Calabi-Yau manifold $M$, and $K_{cs}$ is their K\"{a}hler potential.
We assume that these moduli are stabilized at higher energy scale 
than that of the inflation. Then we can treat $W_0$ and $K_{cs}$ as constants.
The second term in (\ref{W}) is for K\"{a}hler moduli and 
arises from the non-perturbative effect. 
$h^{1,2}$ is the Hodge number of $M$ and equal to the number of 
the K\"{a}hler moduli.
$A_i$ and $a_i$ are model-dependent constants.
(The reason why (\ref{W}) does not have the contribution from $T_1$ 
is described below.)
$\cV$ is the volume of $M$ in the unit of string length scale $l_s$ 
and we assume that $\cV$ has a specific form as
\begin{align}
   \cV & =\alpha\left(\tau_1^{3/2}
   -\sum^{h^{1,2}}_{i=2}\lambda_i\tau_i^{3/2}\right) \nonumber \\ 
   & =\frac{\alpha}{2\sqrt{2}}\left[(T_1+\bar{T}_1)^{3/2}
   -\sum^{h^{1,2}}_{i=2}\lambda_i(T_i+\bar{T}_i)^{3/2}\right], 
   \label{volume}
\end{align} 
where $\alpha$ and $\lambda_i$ are constants which depend on $M$.
The term $\frac{\xi}{2}$ in the logarithm arises from 
$\alpha^{\prime}$-correction~\cite{Becker:2002nn}, 
where $\xi=-\frac{\zeta(3)\chi(M)}{2(2\pi)^3}$ and $\chi(M)$ 
is the Euler-number of $M$.
We assume that $\xi>0$.   
  
Given (\ref{W}) and (\ref{K}), we can derive the potential of 
the K\"{a}hler moduli using the formula
\be
   V=e^{K}(K^{i\bar{j}}D_iW\bar{D}_{\bar{j}}\bar{W} -3|W|^2),
\ee
where $D_i=\partial_i+K_i$, the subscript $i(\bar{i})$ means 
the derivative by $T_i(\bar{T}_i)$ and $K^{i\bar{j}}$ is 
the inverse of the matrix $K_{i\bar{j}}$.
The resultant potential, including the prefactor which 
arises from the conversion from the string frame to 
the Einstein frame, is obtained as\footnote{
We make the factor $e^{K_{cs}}$ absorbed in other constants 
such as $W_0$ and $A_i$.}
\begin{eqnarray}
  \label{general-potential}
  \hspace{-.4cm}{\mbox{\large $V$}}\!\!
  &=&\left(\frac{g_s^4}{8\pi}\right)
  \Bigg[\sum_{\substack{i,j=2 \\ i<j}}^{h^{1,2}}
  \frac{A_iA_j\cos(a_i\theta_i-a_j\theta_j)}
  {(4{\cal{V}}-\xi)(2{\cal{V}}+\xi)^2}
  e^{-(a_i\tau_i+a_j\tau_j)}
  \left(32(2{\cal{V}}+\xi)(a_i\tau_i+a_j\tau_j 
  + 2a_ia_j\tau_i\tau_j)+24\xi\right)\nonumber\\
  \hspace{-.4cm}
  &+&\frac{12W_0^2\xi}{(4{\cal{V}}-\xi)(2{\cal{V}}+\xi)^2}
  +\sum_{i=2}^{h^{1,2}} 
  \left\{ \frac{12e^{-2a_i\tau_i}\xi
  A_i^2}{(4{\cal{V}}-\xi)(2{\cal{V}}+\xi)^2}
  +\frac{16(a_iA_i)^2\sqrt{\tau_i}e^{-2a_i\tau_i}}
  {3\alpha\lambda_i(2{\cal{V}}+\xi)} \right.\\ \nonumber
  \hspace{-.4cm}
  &+&\left. \frac{32e^{-2a_i\tau_i}a_iA_i^2\tau_i(1+a_i\tau_i)}
  {(4{\cal{V}}-\xi)(2{\cal{V}}+\xi)}
  +\frac{8W_0A_ie^{-a_i\tau_i}\cos(a_i\theta_i)}
  {(4{\cal{V}}-\xi)(2{\cal{V}}+\xi)}
  \left(\frac{3\xi}{2{\cal{V}}+\xi}+4a_i\tau_i\right) \right\}\Bigg] ,
\end{eqnarray}
where $g_s$ is the string coupling constant.
Generally, the minimum value of this potential is negative.
In order that the minimum value, i.e., the vacuum energy after 
the inflation, becomes zero,  we have to add the uplifting term, 
for example, the contribution from anti-D3 branes\cite{Kachru:2003aw}.
For the natural values of the parameter,  
$\cV \gg 1$ and $\tau_1 \gg \tau_2,\tau_3,...$ 
at the minimum of (\ref{general-potential}).

That is, the extra dimension is compactified so that its volume 
is extremely large.
Therefore, $e^{-a_1T_1}$ is much smaller than unity and hence 
the term containing this factor is omitted from (\ref{W}).

\begin{figure}[t]
\begin{center}
\includegraphics[width=150mm]{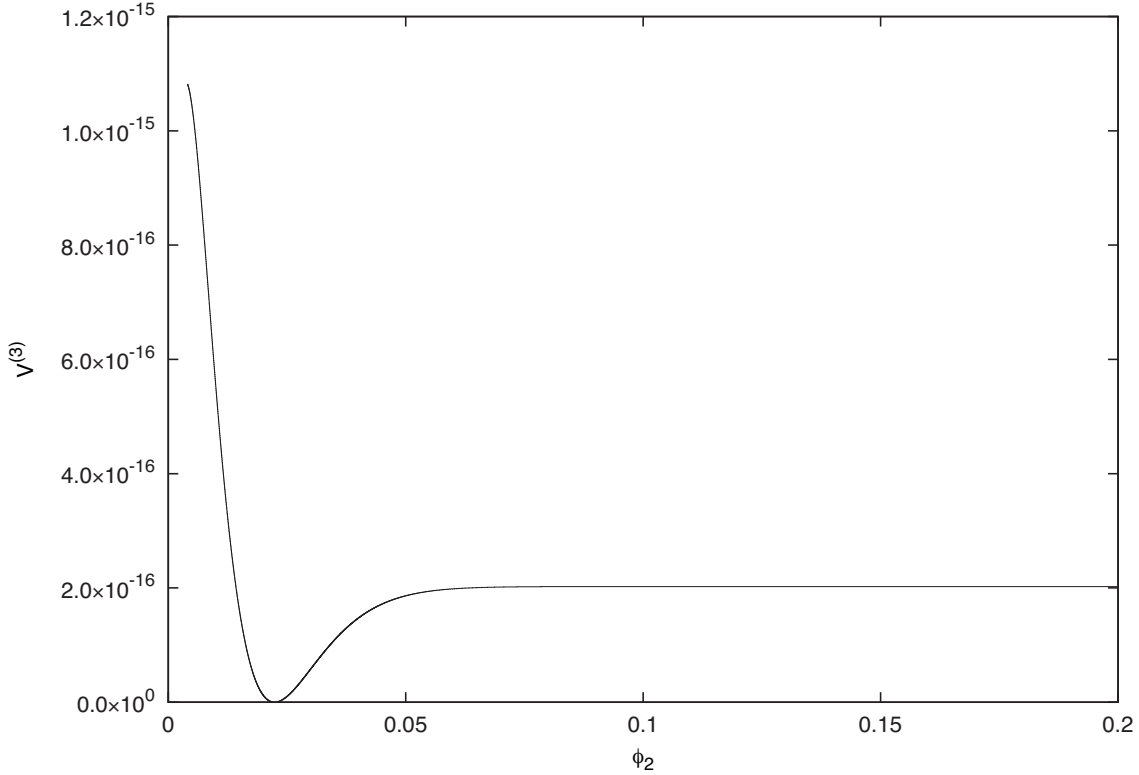}
\caption{The potential of $\phi_2$ in the leading order of 
the expansion into power series of $\cV^{-1}$ under the parameter 
set shown in TABLE 1. The constant term is added such that 
the minimum value is equal to 0.
}
\label{fig:potential}
\end{center}
\end{figure}

$\tau_2,\tau_3,...$ serve for stabilization of $\cV$.
However, if
\be
  \frac{\lambda_n/a_n^{3/2}}{\sum^{h^{1,2}}_{i=2}\lambda_i/a_i^{3/2}} 
  \ll 1
\ee
is satisfied for some $n$, $\tau_n$ can leave its minimum without 
destabilizing $\cV$.
When $\tau_n$ is away from the minimum, the terms containing 
$\tau_n$ in the potential are  extremely small, 
because of the exponential factor $e^{-a_n\tau_n}$.
This make the potential for $\tau_n$ flat enough to realize 
the slow roll inflation\footnote{
The string loop correction to the K\"{a}hler potential might 
spoil the flatness of the potential of K\"{a}hler moduli 
and the fine tuning of the parameters are required in order 
for the inflation to occur\cite{Berg:2005ja,Berg:2007wt,Cicoli:2007xp}.
However, we neglect this subtlety and assume that the flatness 
is maintained.}
.  
From now, we assume that the two real parts $\tau_2,\tau_3$ 
of K\"{a}hler moduli are
inflatons and that all imaginary parts  $\theta_i$ and the other 
real parts are stabilized at their minimum of the potential\footnote{
In fact, the mass of the imaginary part $\theta_i$ is comparable to that of the real part $\tau_i$
and the motion in the direction $\theta_i$ may be as important as that in the direction $\tau_i$
\cite{Bond:2006nc}.
In most inflationary trajectories, both $\tau_i$ and $\theta_i$ evolve at first,
then at some point $\theta_i$ gets trapped at the minimum of the potential 
and after that the field straightly moves in the direction of $\tau_i$.
In addition, this type of inflationary models can easily make the inflation much longer than $50$ or $60$ e-foldings by setting the initial position of the inflaton apart from the minimum of the potential.
Therefore, the situation where all imaginary parts have fallen into the minimum by the last $50$ or $60$ e-foldings is not special and we can assume such a situation.
}
.
In our scenario there has to exist at least four K\"{a}hler moduli
(A concrete example of a Calabi-Yau manifold which has four K\"{a}hler moduli can be found in \cite{Collinucci:2008sq}). 
We treat $\cV$ as a constant since $\tau_1$ is stabilized 
and $\tau_1\gg \tau_2,\tau_3$.
$e^{-a_i\tau_i}\sim\cV^{-1}$ near the minimum, and if we take 
the leading terms in the expansion into power series of $\cV^{-1}$, 
we obtain
\be
  \label{V3rd}
  V^{(3)}=\left(\frac{g_s^4}{8\pi}\right)
  \left(V_0
  -\sum_{i=2,3}\frac{4W_0a_iA_i}{{\cal{V}}^2}\tau_ie^{-a_i\tau_i}
  +\sum_{i=2,3} \frac{8(a_iA_i)^2}{3 \alpha\lambda_i{\cal{V}}}
  \sqrt{\tau_i}e^{-2a_i\tau_i}
  \right)
\ee
where $V_0$ is the part which does not depend on $\tau_i(i\ge2)$, 
including the uplifting term.
It is apparent that the potential for each $\tau_i$ separates from 
the other and the potential takes the form of (\ref{doubleinfVform}). 
During the inflation, when the inflaton is away from the minimum, 
the third term in (\ref{V3rd}) can be also neglected.
The shape of the potential of the single K\"{a}hler moduli is 
depicted in Fig.~\ref{fig:potential}.
This potential is extremely flat for large field value, 
but near the minimum it is much steeper.
This feature of the potential leads to the rapid oscillation 
of the first inflaton about its minimum in the intermediate 
stage of the inflation.

Although we can basically understand the dynamics of the inflaton 
using only the leading part $V^{(3)}$ in the potential, 
there is the subleading part given by
\begin{align}
  V^{(4)}
  =&\left(\frac{g_s^4}{8\pi}\right)
  \Bigg[ -\frac{4A_2A_3}{\cV^2}
  (a_2\tau_2+a_3\tau_3+2a_2a_3\tau_2\tau_3)e^{-a_2\tau_2-a_3\tau_3} 
  \nonumber \\
  &+ \sum_{i=2,3}\left\{
  -\frac{4\xi(a_iA_i)^2\sqrt{\tau_i}}{3\alpha\lambda_i\cV^2}
  e^{-2a_i\tau_i}
  +\frac{4a_iA_i^2\tau_i(1+a_i\tau_i)}{\cV^2}e^{-a_i\tau_i}\right.
  \nonumber \\
  &~~~~~~~~~~\left. -\frac{3\xi W_0A_i}{2\cV^3}e^{-a_i\tau_i}
  +\frac{\xi W_0a_iA_i\tau_i}{\cV^3}e^{-a_i\tau_i}
  \right\}\Bigg].
  \label{V4th}
\end{align}
In this order, the coupling between $\tau_2$ and $\tau_3$ exists.
These term do not affect the 0-th order dynamics of inflatons.
However, the small coupling in this order largely affects
the evolution of the fluctuation of inflatons, as we see below. 
 
The nontrivial form of K\"{a}hler potential (\ref{K}) makes 
the kinetic term for inflatons noncanonical.
The kinetic term is $-K_{i\bar{j}}D_{\mu}T^iD^{\mu}\bar{T}^{\bar{j}}$.
In the leading order of $\cV^{-1}$, $K_{i\bar{j}}$ is given by
\be
   K_{i\bar{j}}\simeq \frac{3\alpha\lambda_i}{8\cV\sqrt{\tau_i}}
   \delta^{ij}. \label{Kij}
\ee
We neglect subleading contributions to $K_{i\bar{j}}$ for simplicity. 
For $K_{i\bar{j}}$ as (\ref{Kij}), we obtain the canonical kinetic 
term by redefinition of fields as follows,
\be
   \phi_i=\sqrt{\frac{4\alpha\lambda_i}{3\cV}}\tau_i^{3/4}.
\ee

\section{Formalism to compute the fluctuation of inflatons}
\label{sec:perturbation_formalism}

Here, we present the formalism~\cite{Salopek:1988qh}
that we adopt to trace the evolution of the fluctuations of 
the inflatons and calculate resultant cosmological perturbations.
 
With use of the Newtonian gauge the perturbed space-time is described
as
\be
 ds^2=-(1-2\Phi(t,\xv))dt^2+a^2(t)(1+2\Phi(t,\xv))d^2\xv,
\ee
where $a(t)$ is the scale factor.
We decompose the values of inflatons into the 0th order homogeneous 
parts and the first order perturbations as 
$\phi_i(t,\xv)=\bar{\phi}_i(t)+\delta\phi_i(t,\xv)$.
Then, the equations which describe the evolution of the scale factor 
and $\bar{\phi}_i$ are
\be
  \ddot{\bar{\phi}}_i+3H\dot{\bar{\phi}}_i+V_i=0, \label{EOMofphi0}
\ee
\be
  H^2=\frac{1}{3}
  \left(\sum_{i=2,3}\frac{1}{2}\dot{\bar{\phi}}_i^2
  + V(\bar{\phi}_2,\bar{\phi}_3)\right),
  \label{Friedmann}
\ee
where a dot denotes time derivative, a subscript $i$ of $V$ means a
derivative of $V$ by $\phi_i$ and $H=\frac{\dot{a}}{a}$ is 
the Hubble parameter.
      
The first order perturbative equations for  
$\Phi$ and $\delta\phi_i$ are written as 
\be
   \delta\ddot{\phi}_i+3H\delta\dot{\phi}_i
   -\frac{1}{a^2}\nabla^2\delta\phi_i+\sum_{j=2,3}V_{ij}\delta\phi_j
   =2V_i\Phi-4\dot{\bar{\phi}}_i\dot{\Phi}, \label{EOMofdphi}
\ee
\be
   \ddot{\Phi}+5H\dot{\Phi}-\frac{1}{3a^2}\nabla^2\Phi
   +\frac{4}{3}V\Phi
   =\frac{1}{3}\sum_{j=2,3}
   \left(2V_j\delta\phi_j-\dot{\bar{\phi}}_j\delta\dot{\phi}_j\right). 
   \label{EOMofPhi}
\ee
Hereafter, we consider the Fourier modes of perturbative quantities,
\be
   \Phi(t,\xv)=\int\frac{d^3k}{(2\pi)^3}
   \left(\Phi(t,\kv)e^{i\kv\cdot\xv}
   +\Phi^{\dagger}(t,\kv)e^{-i\kv\cdot\xv}\right),
\ee
\be
   \delta\phi_i(t,\xv)
   =\int\frac{d^3k}{(2\pi)^3}
   \left( \delta\phi_i(t,\kv)e^{i\kv\cdot\xv}
   + \delta\phi_i^{\dagger}(t,\kv)e^{-i\kv\cdot\xv}\right). 
\ee
Since they have quantum nature, the Fourier modes are written 
in terms of creation and annihilation operators as 
\be
   \delta\phi_i(t,\kv)=\sum_{j=2,3}\psi_{ij}(t,\kv)a_j(\kv), 
   ~~~~~~
   \Phi(t,\kv)=\sum_{j=2,3}f_j(t,\kv)a_j(\kv),
\ee
where $a,a^{\dagger}$ satisfy the following commutation relations:
\be
   [a_i(\kv),a^{\dagger}_j(\kv^{\prime})]
   =(2\pi)^3\delta_{ij}\delta^3(\kv-\kv^{\prime}).
\ee
When the mass-squared matrix $V_{ij}$ has non-diagonal components, 
mixing of fields occurs and each $\delta\phi_i(t,\kv)$ is written 
by the linear combination of various annihilation operators.
In the case under consideration, the non-diagonal components of 
$V_{ij}$ are suppressed by one more factor of $\cV^{-1}$ compared 
with diagonal ones.
As a result, non-diagonal components of $\psi_{ij}$ are also suppressed
unless they are amplified by the resonance effect discussed  
in the next section.
Using this decomposition, (\ref{EOMofdphi}) and (\ref{EOMofPhi}) are
written as
\be
   \psi_{ij}+3H\dot{\psi}_{ij}+\frac{k^2}{a^2}\psi_{ij}
   +\sum_{k=2,3}V_{ik}\psi_{kj}
   =2V_if_j-4\dot{\bar{\phi}}_i\dot{f}_j, 
   \label{EOMofpsi}
\ee
\be
   \ddot{f}_i+5H\dot{f}_i+\frac{k^2}{3a^2}f_i+\frac{4}{3}Vf_i
   =\frac{1}{3}\sum_{j=2,3}
   \left(2V_j\psi_{ji}-\dot{\bar{\phi}}_j\dot{\psi}_{ji}\right). 
   \label{EOMoff}
\ee
The initial condition for $\psi_{ij}$ is given as follows,
\be
  \psi_{ij} \xrightarrow{t\rightarrow 0} 
  \frac{1}{\sqrt{2k}a(t)}
  \exp\left(-ik\int^t\frac{dt^{\prime}}{a(t^{\prime})}\right)\delta_{ij}. 
  \label{iniconofpsi}
\ee
That is, we assume that in the short wavelength limit 
$\frac{k}{a}\rightarrow\infty$, each scalar field approaches 
the massless state in the Minkowski space-time.  
The initial condition for $\Phi$ is given through the Poisson equation as
\be
  \Phi=\frac{a^2}{2k^2}\sum_{j=2,3}
  \left(\dot{\bar{\phi}}_j\delta\dot{\phi}_j
  +3H\dot{\bar{\phi}}_j\delta\phi_j
  +V_j\delta\phi_j\right)
\ee
$\Phi$ should also satisfy the following constraint:
\be
   \dot{\Phi}=-H\Phi
   -\frac{1}{2}\sum_{j=2,3}\dot{\bar{\phi}}_j\delta\phi_j.
\ee
We use this constraint to check the precision of our numerical 
calculation. 
    
The perturbations of the metric and inflatons are related to the 
curvature perturbation as
\be
  \cR=\Phi-\frac{H}{\dot{\sigma}}\delta\sigma,
\ee
where
\be
   \dot{\sigma}=\sqrt{\sum_{j=2,3}\dot{\bar{\phi}}^2_j}, 
   ~~~~~~~
   \delta\sigma=\sum_{j=2,3}\frac{\dot{\bar{\phi}}_j}{\dot{\sigma}}
   \delta\phi_j
\ee
The power spectrum of the curvature perturbation $\cP_{\cR}$, which is 
defined as $\left<|\cR|^2\right>=\int^{\infty}_0\frac{dk}{k}\cP_{\cR}(k)$, 
where $\left< \cdots\right>$ denotes the VEV, is given by
\be
  \cP_{\cR}(k)=\frac{k^3}{2\pi^2}\sum_{j=2,3}
  \left| f_j
  -\frac{H}{\dot{\sigma}^2}\sum_{k=2,3}\dot{\bar{\phi}}_k\psi_{kj}
  \right|^2
  \label{P_Rformula}
\ee

\section{Power spectrum of the curvature perturbation}
\label{sec:powerspectrum}
 
Given the scalar potential in Sec.~\ref{sec:Kahler_moduli} 
and the formalism to compute 
the perturbations in Sec.~\ref{sec:perturbation_formalism}, 
we now proceed to calculate 
the power spectrum of the curvature perturbation.
In the following, we use the parameter set shown in Table 1.
Under this parameter set, the global minimum of the potential is 
located at
\be
 \phi_{2,min}=2.24\times10^{-2}, 
 ~~~~~~~~
 \phi_{3,min}=3.06\times10^{-2}
\ee

\begin{table}[t]
\begin{center}
 \begin{tabular}{|c|c|c|c|c|c|c|c|c|c|c|}
 \hline
 $\cV$ & $W_0$ & $a_2$ & $a_3$ 
 & $A_2$ & $A_3$ & $\lambda_2$ 
 & $\lambda_3$ & $\alpha$ & $\xi$ & $g_s$ \\
 \hline
 $2.45\times10^{4}$ & $1.81$ & $\frac{2\pi}{30}$ & $\frac{2\pi}{5}$ 
 & $1.13\times10^{-3}$ & $0.452$ & $0.3$ 
 & $1.0$ & $1.0$ & $0.5$ & $0.1$\\
 \hline
 \end{tabular}
 \caption{the parameter set used for the numerical calculation.}
 \label{table:parameters}
\end{center}
\end{table}
 
\begin{figure}[htbp]
\begin{center}
	\includegraphics[width=130mm]{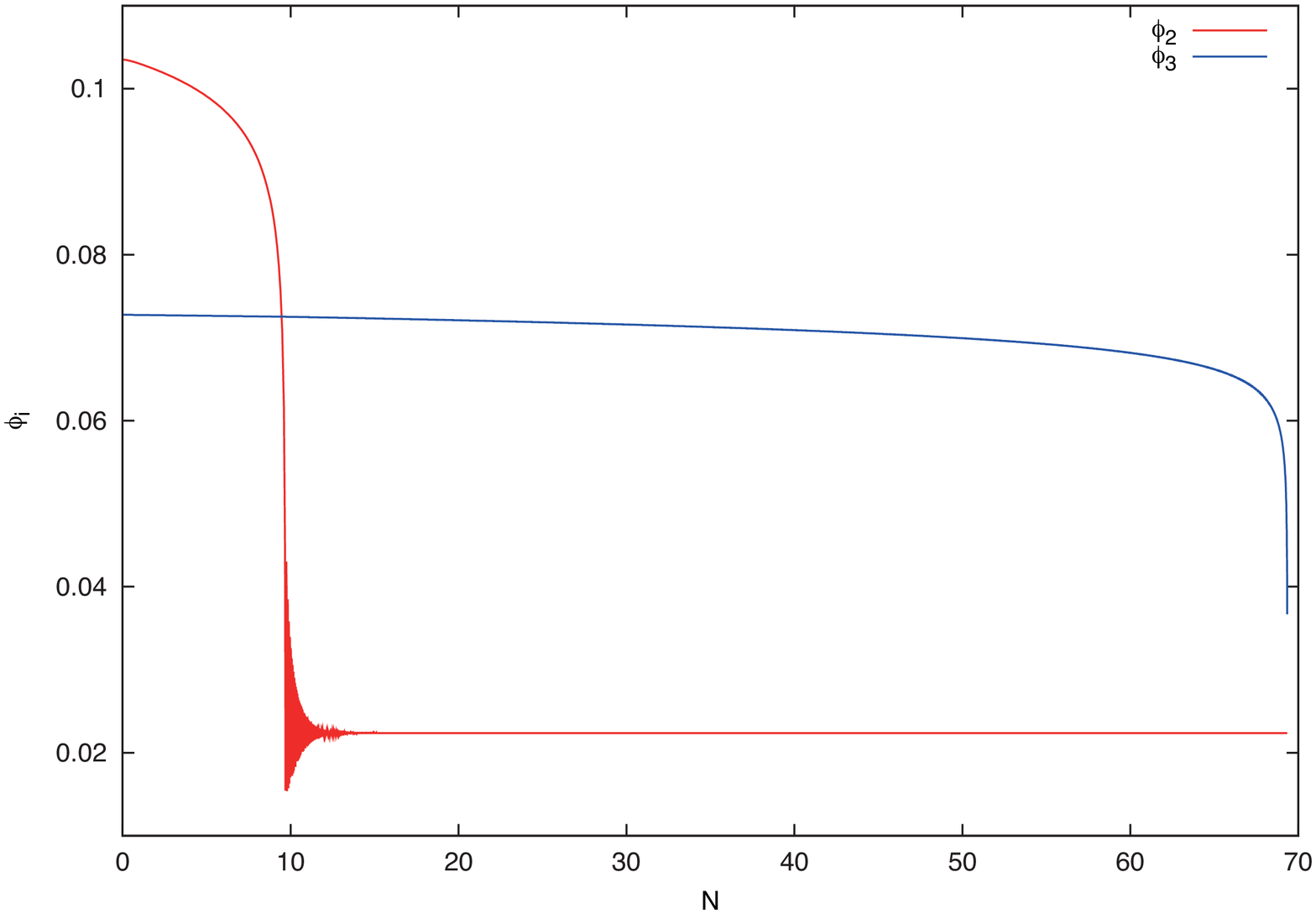}
 	\includegraphics[width=130mm]{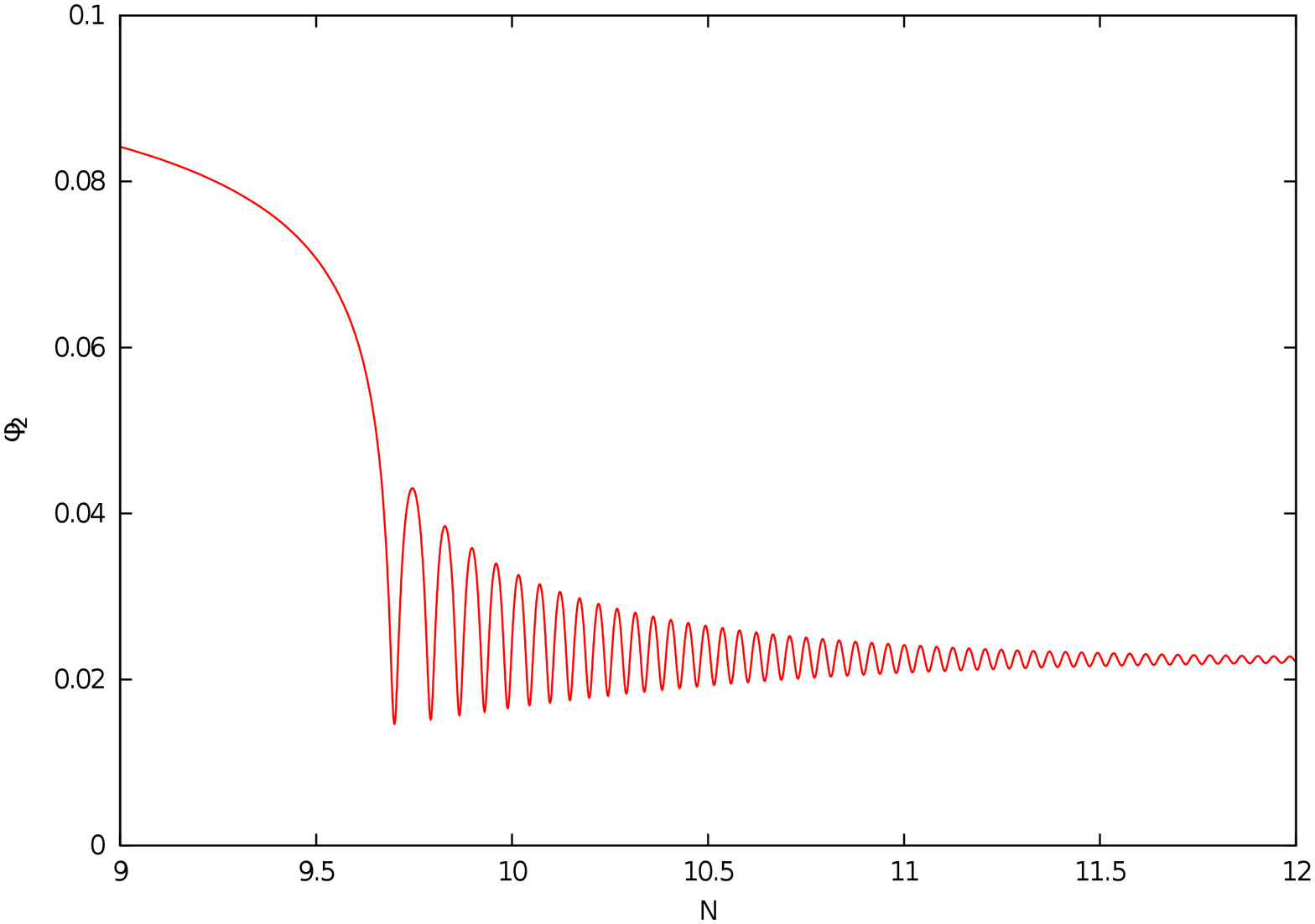}
  \caption{Evolution of homogeneous parts of inflatons. 
  The red line corresponds to $ \bar{\phi}_2$ and blue one to 
  $\bar{\phi}_3$.  The upper panel shows the motion of inflatons 
  throughout 
  the inflation and the lower panel focuses on the rapid oscillation 
  of $\phi_2$ after it falls down to its minimum.
  }
  \label{fig:scalarevolve0th}
\end{center}
\end{figure}

\subsection{Evolution of homegeneous parts}
 
The evolution of the homogeneous parts of inflatons can be 
traced by numerically integrating (\ref{EOMofphi0}).
We set the initial values of inflatons as
\be
 \tau_{2,ini}=75.5, ~~~~~~~~\tau_{3,ini}=21.15,
\ee
and those of time derivatives of inflatons to zero.
As in most double inflation models, we have to tune the initial position
of the second inflaton $\tau_3$ such that the second stage of the inflation is not too long
and the signal from the first stage remains in the observable region. 
The evolution until $\phi_3$ falls into its minimum is described 
in Fig.\ref{fig:scalarevolve0th}.
The horizontal axis denotes the e-folding number counted from the start 
of the evolution.
We find that two inflaton actually evolve independently.
After the first stage of the inflation $\bar{\phi}_2$ oscillates 
about its minimum and its amplitude decreases as the universe 
expands.
Then, the potential energy of $\phi_3$ dominates the universe and 
the second inflation takes place.

The energy scale of the inflation in the first stage and the second 
stage are, $V^{1/4}_1\simeq  1.60\times10^{-4}$, 
$V^{1/4}_2\simeq 1.46\times 10^{-4}$, respectively.
Therefore, no significant gravitational waves  cannot be generated 
in this  model.

\subsection{Evolution of perturvations}

\begin{figure}[t]
\begin{center}
  \includegraphics[width=130mm]{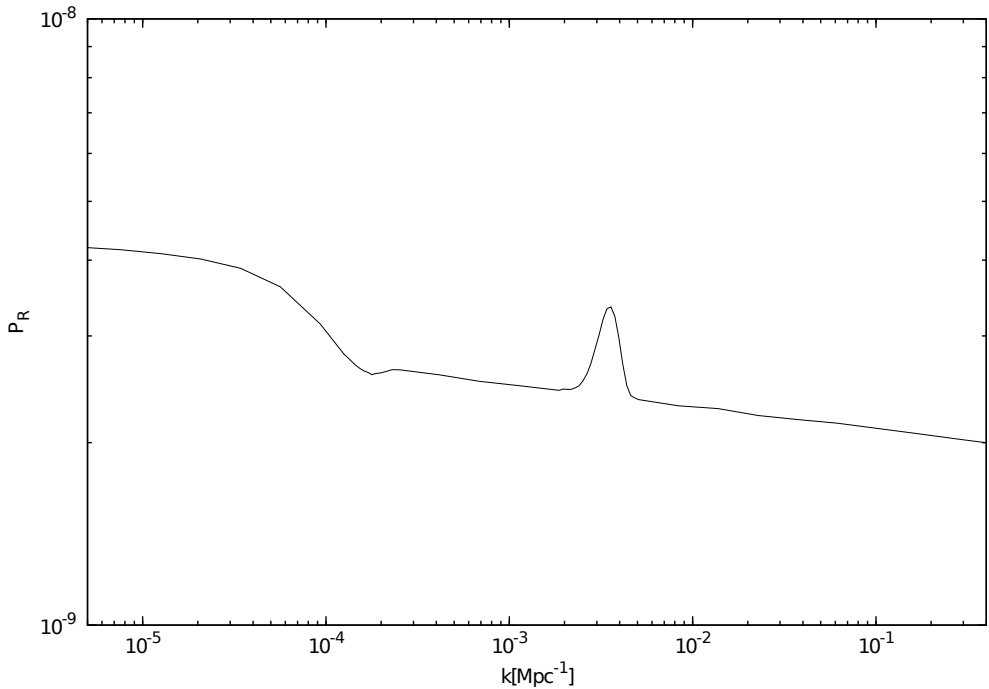}
  \caption{The power spectrum of the curvature perturbation 
  generated by the inflation model in this paper.
  The horizontal line denotes the wavenumber at present.}
  \label{fig:P_R}
\end{center}
\end{figure}

Then, let us see how cosmological perturbations evolve and 
how the power spectrum is generated by the inflation.
They are found by numerically integrating (\ref{EOMofpsi}) 
and (\ref{EOMoff}) until the Fourier modes of interest exit 
the horizon and using the formula (\ref{P_Rformula}).
The resultant power spectrum of the curvature perturbation is 
shown in Fig.~\ref{fig:P_R}.
We assume that the mode whose wavenumber is equal to $k=0.002Mpc^{-1}$ 
at present exits the horizon at the time when the e-folding number 
counted backward from the end of the inflation is $N=56$.\footnote{
We assume that the peak of the power spectrum is located at the scale 
which is of observational interest as discussed in the next section.
}

In the spectrum one can see the existence of a gentle step at
$k \sim 10^{-4}$~Mpc$^{-1}$. 
The perturbations with wavenumber larger than $10^{-4}$~Mpc$^{-1}$
exit the horizon during the first inflation, 
while thoese with smaller wavenumber exit during the second 
inflationary stage. Since the Hubble parameter is larger 
and both inflatons contribute to the curvature 
perturbation in the first inlationary stage, the amplitude of the
perturbation generated during the first inflation is larger, which 
makes the gentle step at $k \sim 10^{-4}$~Mpc$^{-1}$. 

In addition to the step in the spectrum, there is a small peak 
at $k\simeq 3\times 10^{-3}$~Mpc$^{-1}$.
Except these points, the shape is same as the power spectrum 
generated by the usual slow-roll inflation, which is nearly 
scale-invariant and can be approximated by power law, 
$\cP_{\cR}(k)\propto k^{n_s-1}$.
If we neglect the peak, the amplitude and the spectrum index are
\be  
   \cP_{\cR}\simeq 2.43\times10^{-9}, ~~~~~~\ n_s\simeq0.965
\ee
at $k=0.002Mpc^{-1}$ which are consistent with WMAP 7-year 
data~\cite{Larson:2010gs}.

\begin{figure}[ht]
\begin{center}
  \includegraphics[width=120mm]{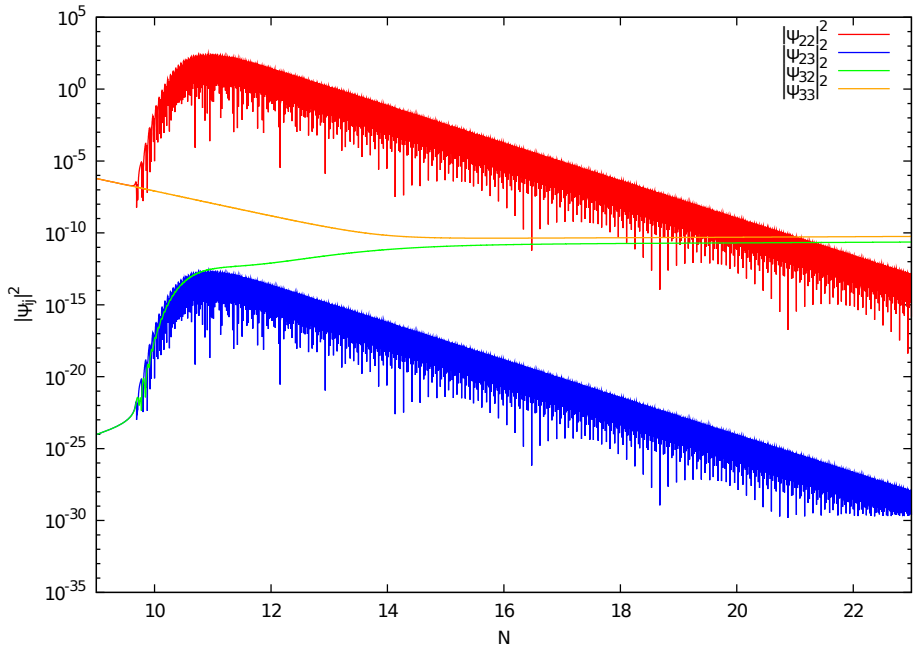}
  \caption{The evolution of $|\psi_{ij}|^2$.
  The red, blue, green, orange lines correspond to 
  $\psi_{22},\psi_{23},\psi_{32},\psi_{33}$ respectively.
  The horizontal line denotes the e-folding number counted from 
  the beginning of the inflation.}
  \label{fig:psievolve}
\end{center}
\end{figure}

\begin{figure}[p]
\begin{center}
   \includegraphics[width=120mm]{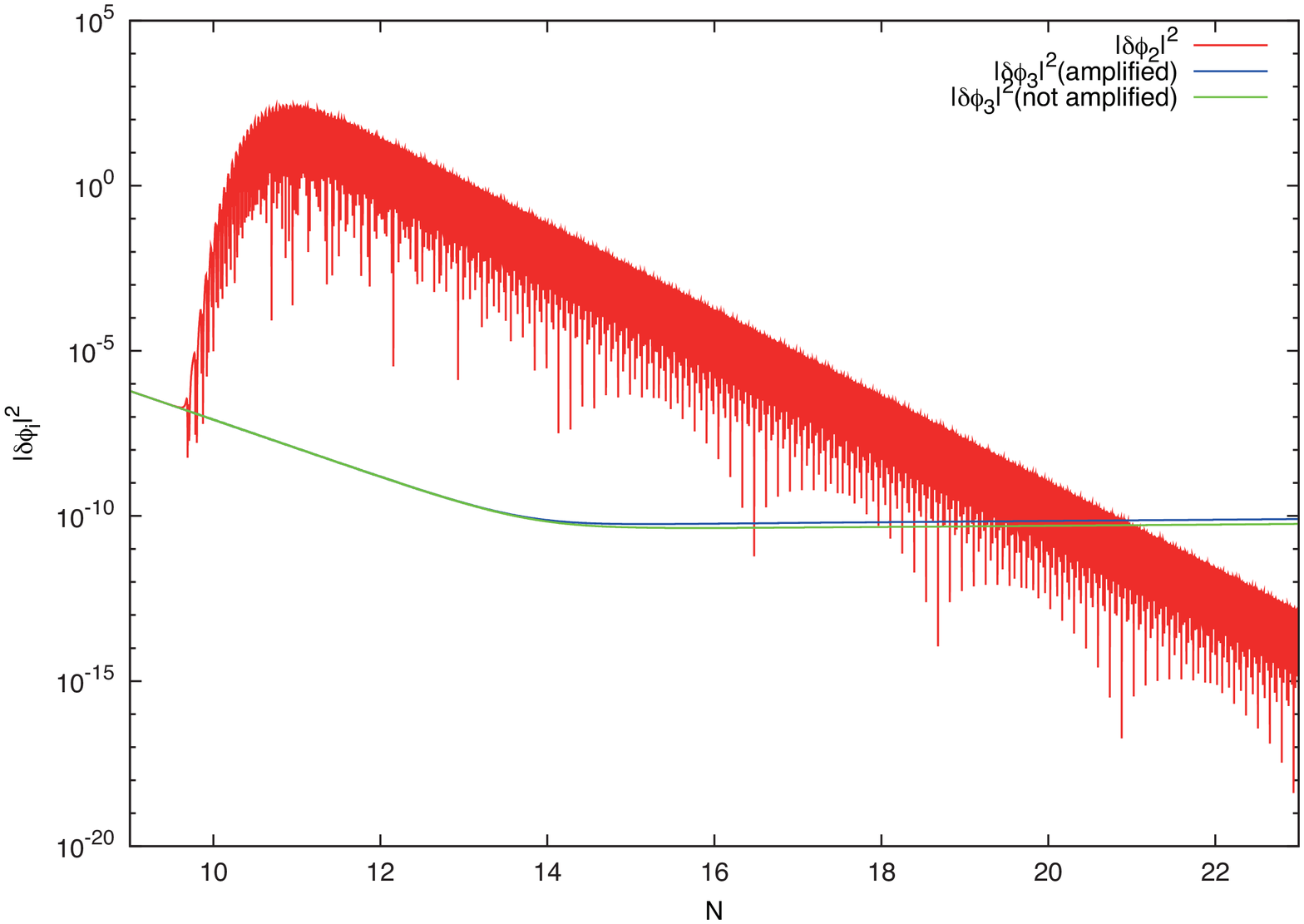}
   
   \includegraphics[width=120mm]{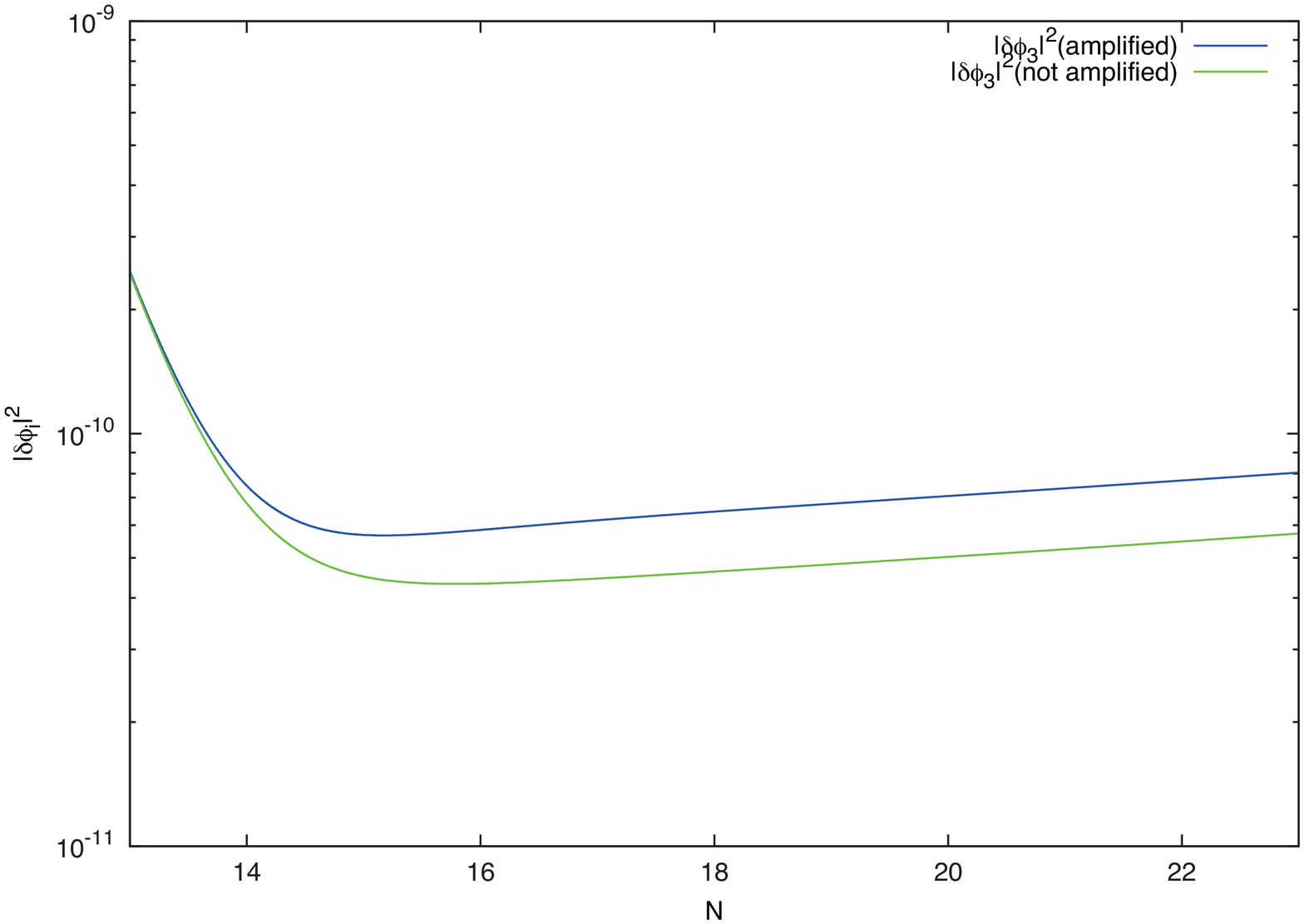}
   \caption{The upper panel shows the evolution of 
   $\left<|\delta\phi_i|^2\right>=\sum_j|\psi_{ij}|^2$.
   The red and blue lines correspond to $\delta\phi_2$ and $\delta\phi_3$ 
   respectively.
   The green line represents the evolution of 
   $\left<|\delta\phi_3|^2\right>$ without the amplification by 
   the resonance.
   The lower panel focus on the difference between the ways 
   for $\left<|\delta\phi_3|^2\right>$ to
   evolve with and without the amplification. 
   The horizontal line denotes the e-folding number counted from 
   the beginning of the inflation.}
   \label{fig:delphievolve}
\end{center}
\end{figure}

Then, let us consider the physics behind the peak in the power spectrum.
In order to understand it, we need to see how the fluctuations of 
the two inflatons evolve.
The evolution of $|\psi_{ij}|^2$ and 
$\left<|\delta\phi_i|^2\right>=\sum_j|\psi_{ij}|^2$
are shown in Fig.~\ref{fig:psievolve} 
and Fig.~\ref{fig:delphievolve}, respectively.

Before the onset of the oscillation of $\phi_2$, both $\delta\phi_2$ 
and $\delta\phi_3$ are fluctuations of nearly massless scalar fields, 
which can be approximated by  (\ref{iniconofpsi}).
This is why the evolutions of $\psi_{22}$ and $\psi_{33}$, 
or $\psi_{23}$ and $\psi_{32}$ are identical before $N\simeq10$. 
However, $\psi_{22},\psi_{23},\psi_{32}$ are exponentially enhanced 
for $10\lesssim N\lesssim 11$, while $\psi_{33}$ continues decreasing.
The period during which they are enhanced depends on how the amplitude 
of $\phi_2$ decreases. 
Ignoring metric perturbations, the evolution equations of $\psi_{ij}$
are
\be
   \ddot{ \psi}_{22}+3H\dot{\psi}_{22}+\frac{k^2}{a^2}\psi_{22}
   +V_{22}(\phi_2,\phi_3)\psi_{22}+V_{23}(\phi_2,\phi_3)\psi_{32} 
   =0, 
   \label{EOMofpsi22}
\ee
\be
   \ddot{\psi}_{23}+3H\dot{\psi}_{23}+\frac{k^2}{a^2}\psi_{23}
   +V_{22}(\phi_2,\phi_3)\psi_{23}+V_{23}(\phi_2,\phi_3)\psi_{33}
   =0,
   \label{EOMofpsi23}
\ee
\be
   \ddot{\psi}_{32}+3H\dot{\psi}_{32}+\frac{k^2}{a^2}\psi_{32}
   +V_{32}(\phi_2,\phi_3)\psi_{22}+V_{33}(\phi_2,\phi_3)\psi_{32}
   =0.
   \label{EOMofpsi32}
\ee
\be
   \ddot{\psi}_{33}+3H\dot{\psi}_{33}+\frac{k^2}{a^2}\psi_{33}
   +V_{32}(\phi_2,\phi_3)\psi_{23}+V_{33}(\phi_2,\phi_3)\psi_{33}
   =0.
   \label{EOMofpsi33}
\ee
Remembering that the diagonal components of $V_{ij}$ are greater 
than the non-diagonal ones by a factor $\cV$ and so are those 
of $\psi_{ij}$, the dominant term in (\ref{EOMofpsi22}) is the fourth term.
Since $\phi_2$ oscillation makes $V_{22}$ oscillate  
$\psi_{22}$ has the oscillating mass term.
This leads to the exponential growth of $\psi_{22}$ by the parametric 
resonance.
The reason why $\psi_{23}$ grows is similar and the responsible term 
in (\ref{EOMofpsi23}) is fourth one.
The rapid growth of the fluctuation of an inflaton due to parametric 
resonance in single K\"{a}hler moduli inflation was pointed out 
in~\cite{Barnaby:2009wr} where  the possibility that the backreaction by 
the created particle by the resonance immediately becomes important was
also pointed out.
If the energy stored in the amplified perturbation exceeds or is 
comparable to that of background condensation 
$V(\bar{\phi}_2,\bar{\phi}_3)$, we have to take into account 
the backreaction seriously.
However, under the parameter set of Table~\ref{table:parameters}, 
we find that the energy of the perturbation does not exceeds 
$V(\bar{\phi}_2,\bar{\phi}_3)$, which allows us to ignore the 
backreaction.
As the amplitude of the oscillation of $\bar{\phi}_2$ decreases, 
the parameteric resonance becomes inefficient and eventually 
the growth of $\psi_{22}$ and $\psi_{23}$ stops.
After that, their amplitudes  decrease by the cosmic expansion.
The total fluctuation of $\phi_2$, 
$\left<|\delta\phi_2|^2\right>=\sum_j|\psi_{2j}|^2$ shows similar 
behavior as seen in Fig.~\ref{fig:delphievolve}, that is, it once 
grows and finally decreases.
Therefore, it does not contribute to the final value of the curvature 
perturbation which exits the horizon in the second stage of the inflation.
 
$\psi_{32}$ grows in a different way from $\psi_{22}$ and $\psi_{23}$.
Because of the exponential growth of $\psi_{22}$, 
(\ref{EOMofpsi32}) is dominated by the fourth term, $V_{32}\psi_{22}$.
Then, this term behaves as the source term and $\psi_{32}$ 
experiences the forced oscillation, which amplifies $\psi_{32}$.
$\psi_{32}$ continues growing even after $\psi_{22}$ stops growing, 
since the amplitude of $\psi_{22}$ is still large.
The situation that the fluctuation of the second inflaton is amplified 
through forced oscillation by that of the first inflaton enhanced by 
the resonance effect is similar to that in~\cite{Kawasaki:2006zv}.
Under the present parameters, $\psi_{32}$, which is originally 
suppressed compared with $\psi_{33}$, is amplified as large as
$\psi_{33}$.\footnote{
Since there are neither a largely oscillating mass term 
nor a source term which induces forced oscillation in the equation 
of motion of $\psi_{33}$, it shows same behavior as in the case 
of the usual slow-roll inflation, that is, it decreases until the mode 
exits the horizon and after that it remains constant. }
As a result, the total fluctuation of $\phi_3$, 
$\left<|\delta\phi_3|^2\right>=\sum_j|\psi_{3j}|^2$ is amplified
by a $\mathcal{O}(1)$ factor compared with that in the situation 
without above effects, as shown in Fig.~\ref{fig:delphievolve}.
Since this is the only scalar field fluctuation which survives in the 
second stage of the inflation, it determins the amplitude of the curvature 
perturbation which exits the horizon in the second stage.
The amplification of the curvature perturbation is also $\mathcal{O}(1)$.

The Fourier modes which experiences amplification are those 
which exist in the resonance band while $\bar{\phi}_2$ is 
rapidly oscillating and are somewhat subhorizon.
Modes whose wavelength are comparable to or exceed the horizon scale, 
or much shorter than it are not affected.
Therefore the spectrum has a sharp peak as shown in Fig.~\ref{fig:P_R}.

In this paper, we choose parameters so that the amplification of 
perturbation does not become too large in order to neglect the 
backreaction and discuss the effect of the above power spectrum 
on the CMB spectrum in the next section.

\section{TT-spectrum of CMB based on the power 
spectrum with small peak}
\label{sec:TTspectrum}

\begin{figure}[t]
\begin{center}
  \includegraphics[width=120mm]{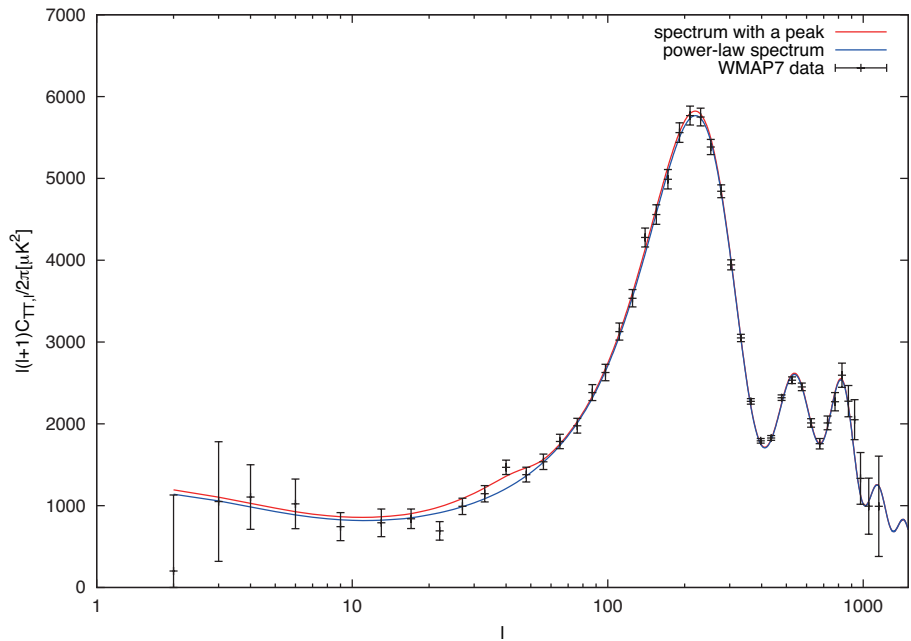}
  \caption{TT-spectra of the CMB.
  Red one represents the TT-spectrum calculated by CAMB using 
  the power spectrum of Fig.~\ref{fig:P_R}.
  Blue one is the spectrum based on the power-law power spectrum 
  whose amplitude and slope are equal to WMAP-7year best-fit values.
  Black points are the observed values of WMAP-7year and error bars 
  denote the sum of cosmic variance and systematic errors.}
  \label{fig:C_TT}
\end{center}
\end{figure}

Standard slow-roll inflation models predict a nearly scale-independent 
power spectrum, which is consistent with various cosmological 
observations, in particular measurements of CMB anisotropies
such as WMAP~\cite{Larson:2010gs}.
However, several data points deviate from the predicted spectrum 
which is based on a simple power-law power spectrum, 
$\cP_{\cR}\propto k^{n_s-1}$.
Although such deviations is not statistically significant enough, 
there is a possibility that they indicate the feature in the power 
spectrum which cannot be described by power-law.
For example, there is a upward deviation of the observational value 
around $\ell \simeq 40$.
This might suggest that there is a small peak of the power spectrum around the scale corresponding to 
$\ell \simeq 40$~\cite{Hoi:2007sf,Barnaby:2009dd}.
In the previous section we have seen that the moduli double 
inflation predicts a sharp peak in the power spectrum. 
In fact, we can choose the initial value of the second inflaton so that 
the peak is located around such scale.
Besides, we can tune the $\mathcal{V}$, which determines the strength of the coupling 
between $\tau_2$ and $\tau_3$, in order to get the desired height of the peak. 
In Fig.~\ref{fig:C_TT} we show the TT power spectrum calculated using 
the spectrum of Fig.~\ref{fig:P_R} and the computer code 
{\tt CAMB}~\cite{CAMB}.
As for the cosmological parameters other than thoes related 
power spectrum, we use the WMAP-7year best-fit 
values~\cite{Komatsu:2010fb}.
The TT spectrum base on the simple power-law power spectrum and 
best-fit parameters of WMAP-7year is also shown in Fig.~\ref{fig:C_TT}.

The resultant TT-spectrum actually has a small bump around 
$\ell\simeq 40$, which fits the data points better than that 
based on the simple power-law power spectrum. 
We can calculate the likelihood of each TT-spectrum using the 
code which is available at the website of LAMBDA~\cite{LAMBDA}.
We get $-2\ln\mathcal{L}\simeq1236$ for that based on the power-law 
power spectrum, and $-2\ln\mathcal{L}\simeq1233$ for that based on 
the spectrum of Fig.~\ref{fig:P_R}.
Thus, there is actually slight improvement of the fit, 
$\Delta(-2\ln\mathcal{L})\simeq 3$.
Note that in the model under consideration there are two more parameters 
which can be tuned to improve the fit to WMAP than the simplest slow-roll inflation model,
that is, the initial position of $\tau_3$ and $\mathcal{V}$.    

\section{Summary}
\label{sec:summary}

In this paper, we have considered K\"{a}hler moduli inflation and
pointed out that the double inflation takes place when we introduce 
two K\"{a}hler moduli in the inflationary dynamics.
Because of plethora of moduli in the general string compactification 
scenario, the double inflation or the inflation which consists of many stages 
is natural rather than possible.
In the K\"{a}hler moduli inflation, the small coupling between the
two K\"{a}hler moduli is automatically introduced.
We have found that through this coupling, the oscillation of 
the first inflaton in the intermediate period between the two 
inflationary stages amplifies the fluctuation of the second inflaton 
and that this results in the peak of the power spectrum of 
the curvature perturbation.
It is interesting that even if all Fourier modes of observational 
interest exit the horizon in the final stage of the inflation, 
the information of the previous iflation can be left in the observable 
power spectrum.
We have numerically calculated the power spectrum with a peak 
under a specific parameter set and applied it to TT-spectrum of the CMB.
We have seen that the peak makes the TT-spectrum better fitted 
to the data of WMAP-7year, which has an upward deviation 
from the prediction of the simple power-low spectrum 
around $\ell\simeq 40$.


\section*{Acknowledgment}
We would like to thank Toyokazu Sekiguchi, Sachiko Kuroyanagi, 
Motohiko Kusakabe and Fuminobu Takahashi for many useful discussions.
This work is supported by Grant-in-Aid for Scientific research from 
the Ministry of Education,
Science, Sports, and Culture (MEXT), Japan, No.14102004  
and No. 21111006 and also by World Premier International
Research Center Initiative (WPI Initiative), MEXT, Japan.


 \end{document}